\documentclass[10pt,conference]{IEEEtran}
\IEEEoverridecommandlockouts

\usepackage{cite}
\usepackage{todonotes}
\usepackage{amsmath,amssymb,amsfonts}
\usepackage{algorithmic}
\usepackage{graphicx}
\usepackage{textcomp}
\usepackage{xcolor}
\usepackage{url}
\usepackage{float}
\usepackage{algorithm}
\usepackage{algorithmic}
\newcounter{pseudocode}
\renewcommand{\thepseudocode}{Pseudocode \arabic{pseudocode}}

\usepackage{orcidlink}
\usepackage{subcaption}
\usepackage[moderate, tracking=normal, title=normal]{savetrees}

\newcommand\algorithmicprocedure{\textbf{procedure}}
\newcommand{\algorithmicendprocedure}{\algorithmicend\ \algorithmicprocedure}
\makeatletter
\newcommand\PROCEDURE[3][default]{%
  \ALC@it
  \algorithmicprocedure\ \textsc{#2}(#3)%
  \ALC@com{#1}%
  \begin{ALC@prc}%
}
\newcommand\ENDPROCEDURE{%
  \end{ALC@prc}%
  \ifthenelse{\boolean{ALC@noend}}{}{%
    \ALC@it\algorithmicendprocedure
  }%
}
\newenvironment{ALC@prc}{\begin{ALC@g}}{\end{ALC@g}}
\makeatother

\begin{document}

\title{LoCoML: A Framework for Real-World ML Inference Pipelines}


\author{\IEEEauthorblockN{Kritin Maddireddy\textsuperscript{\textdagger} \orcidlink{0009-0003-0720-601X}, Santhosh Kotekal Methukula\textsuperscript{\textdagger} \orcidlink{0009-0001-6775-5059}, Chandrasekar Sridhar\textsuperscript{\textdagger}~\orcidlink{0009-0009-3679-6331}, Karthik Vaidhyanathan\textsuperscript{\textdagger}~\orcidlink{0000-0003-2317-6175}}
\IEEEauthorblockA{\textit{Software Engineering Research Center},  \textit{IIIT Hyderabad}, India\\
kritin.maddireddy@students.iiit.ac.in,
santhosh.km@students.iiit.ac.in, chandrasekar.s@research.iiit.ac.in,\\
karthik.vaidhyanathan@iiit.ac.in
}
\thanks{\textsuperscript{\textdagger} These authors contributed equally to this work.}
}


\maketitle

\begin{abstract}
The widespread adoption of machine learning (ML) has brought forth diverse models with varying architectures, data requirements, introducing new challenges in integrating these systems into real-world applications. Traditional solutions often struggle to manage the complexities of connecting heterogeneous models, especially when dealing with varied technical specifications. These limitations are amplified in large-scale, collaborative projects where stakeholders contribute models with different technical specifications. 
To address these challenges, we developed LoCoML, a low-code framework designed to simplify the integration of diverse ML models within the context of the \textit{Bhashini Project} - a large-scale initiative aimed at integrating AI-driven language technologies such as automatic speech recognition, machine translation, text-to-speech, and optical character recognition to support seamless communication across more than 20 languages. Initial evaluations show that LoCoML adds only a small amount of computational load, making it efficient and effective for large-scale ML integration. Our practical insights show that a low-code approach can be a practical solution for connecting multiple ML models in a collaborative environment.
\end{abstract}


\section{Introduction}
\label{sec:intro}
Integrating machine learning (ML) systems into complex, real-world applications brings engineering challenges that go beyond traditional software engineering challenges~\cite{Dagstuhl,microsoftSEforML}. ML-based systems require continuous data management, frequent model updates, and robust workflows to link various ML components, which are often sourced from different providers with varying technical  compatibility~\cite{muccini2021software,brambilla2017model,mde4mlslr,mlopsCollaboration}. These challenges are especially significant in large-scale, collaborative projects. We were faced with one such challenge in the \textit{Bhashini Project}~\footnote{https://bhashini.gov.in/}, which had multiple academic/industrial partners contribute models with distinct architectures and formats towards building a nationwide AI platform. Ensuring that different models work smoothly as part of a unified, high-quality system demands considerable engineering effort~\cite{bock2021low,cabot2020positioning,TrinityCollaboration}.

\begin{figure}[htbp]
\centerline{\includegraphics[width=\columnwidth]{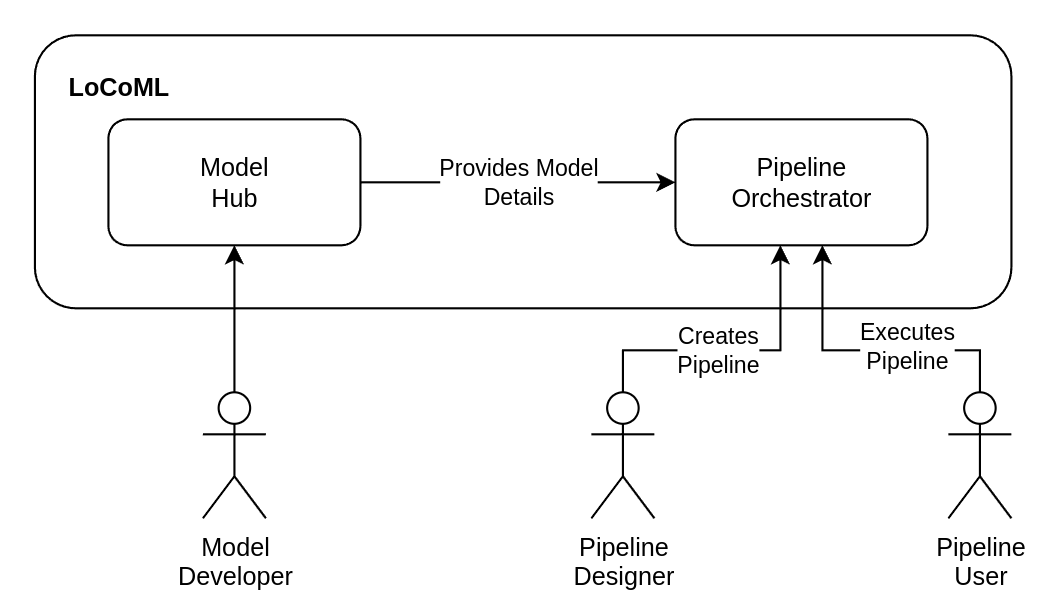}}
    \caption{Overview of LoCoML Framework}
    \label{fig:locoml_approach}
\end{figure}

While some ML platforms provide pipeline management and model deployment tools, they often lack the flexibility required for projects with the scale and diversity of \textit{Bhashini Project}. Many existing platforms also require high levels of coding expertise, creating a barrier for teams with varied technical backgrounds~\cite{TrinityCollaboration}. This restricts accessibility and slows development, as users must rely on expert developers for even minor adjustments. In the \textit{Bhashini Project}, the need to coordinate multiple, heterogeneous models to achieve accurate, aligned inferences has highlighted gaps in current frameworks, necessitating a customized solution. Many existing solutions impose limitations in compatibility and scalability, making it challenging to support workflows where independently developed models must work together within a single pipeline~\cite{lewis2021software,mde4mlslr,mlopsCollaboration}. 

The need for such a flexible, accessible solution is particularly evident in workflows involving a series of interdependent models, such as speech-to-text processing. In this domain, one model might transcribe speech to text, another handles language translation, and further models manage additional language-specific nuances. Integrating these models into a single, reliable pipeline that ensures high-quality output at each stage is a complex engineering task, further complicated by the lack of frameworks that support seamless integration across diverse models with specialized functions. To address these real-world demands, LoCoML was developed as a low-code ML engineering framework, providing flexibility and modularity in creating and managing ML inference pipelines. To address the challenges of managing and orchestrating ML models in complex workflows, LoCoML (As in Figure~\ref{fig:locoml_approach}) is organized around two main components: the Model Hub and the Pipeline Orchestrator. Together, these parts handle everything from storing and retrieving models to executing pipelines, with different user roles involved to keep things running smoothly.



Inspired by model-driven engineering (MDE)~\cite{schmidt2006model} practices and supported by low-code principles, LoCoML abstracts technical complexities, allowing users to focus on the application logic rather than the underlying engineering details. Additionally, the framework’s low-code design invites users with diverse technical backgrounds to contribute to ML workflows, improving accessibility and reducing the time and expertise required to develop robust ML systems~\cite{brambilla2017model,di2020democratizing,TrinityCollaboration}. 

This paper presents LoCoML’s role within the \textit{Bhashini Project}, demonstrating practical solutions for the complex requirements of large-scale, multi-component ML systems and providing insights that can inform the development of similar applications. Through this experience, we aim to share actionable strategies for addressing complex ML engineering challenges and illustrate the effectiveness of MDE and low-code approaches in demanding environments. 
The remainder of the paper is structured as follows. Section~\ref{sec:motivation} 
provides background on \textit{Bhashini Project} and explains its role as a case study.
Section~\ref{sec:locoml} describes the LoCoML framework and its core components. Section~\ref{sec:results}, presents the preliminary results achieved with the framework. 
A review of related work is in Section~\ref{sec:related_work}, and Section~\ref{sec:conclusion} concludes with summary and future research directions.

\section{Bhashini Project}
\label{sec:motivation}

The \textit{Bhashini Project} is a large-scale initiative focused on breaking down language barriers by enabling digital services in multiple languages. The project integrates various AI-driven language technologies, including Text-to-Speech (TTS), Automatic Speech Recognition (ASR), Machine Translation (MT), and Optical Character Recognition (OCR). By combining these tools, \textit{Bhashini Project} facilitates seamless communication across more than 20 languages, providing a unified platform that allows people to access and interact with content in their preferred language.

A range of stakeholders support this initiative, contributing resources, expertise, and data to expand its capabilities. Small and medium enterprises (SMEs) and private organizations with substantial digital reach offer technical support and data to enhance the project’s language resources. Additionally, local language organizations and individual users contribute through a crowdsourcing platform, enriching the language data and making the platform more representative. Together, these efforts create a collaborative system that addresses the diverse linguistic needs of the public and supports the integration of advanced language models into real-world applications.

The diversity and scale of the \textit{Bhashini Project} created unique challenges in integrating multiple AI-driven language models into a single cohesive system. The need to coordinate between different model architectures, manage dynamic data flows, and ensure high-quality results across various languages required a flexible and robust framework that could adapt to evolving requirements. Motivated by these challenges, we developed LoCoML to streamline the integration and management of these diverse models. LoCoML’s low-code design and modular approach address the complexity of combining technologies like TTS, ASR, MT, and OCR, making it possible for the \textit{Bhashini Project} to deliver consistent, reliable, and scalable language services. This framework enables seamless collaboration across stakeholders, allowing them to contribute and refine models effectively while enhancing the platform’s accessibility and usability for a broader audience.

\section{LoCoML Approach}
\label{sec:locoml}
ed in Figure~\ref{fig:locoml_approach}, LoCoML co themprises system two primary subsystems: the \textit{Model Hub} and the \textit{Pipeline Orchestrator}. Within this system, we have identified three distinct user roles. The first is the \textit{Model Developer}, responsible for creating the models utilized within the pipelines, which are stored in and retrieved from the \textit{Model Hub}. The other two roles, the \textit{Pipeline Designer} and the \textit{Pipeline User}, engage directly with the \textit{Pipeline Orchestrator}.

\subsection{The Model Hub}

\begin{figure}[htbp]
    \centerline{\includegraphics[width=\columnwidth]{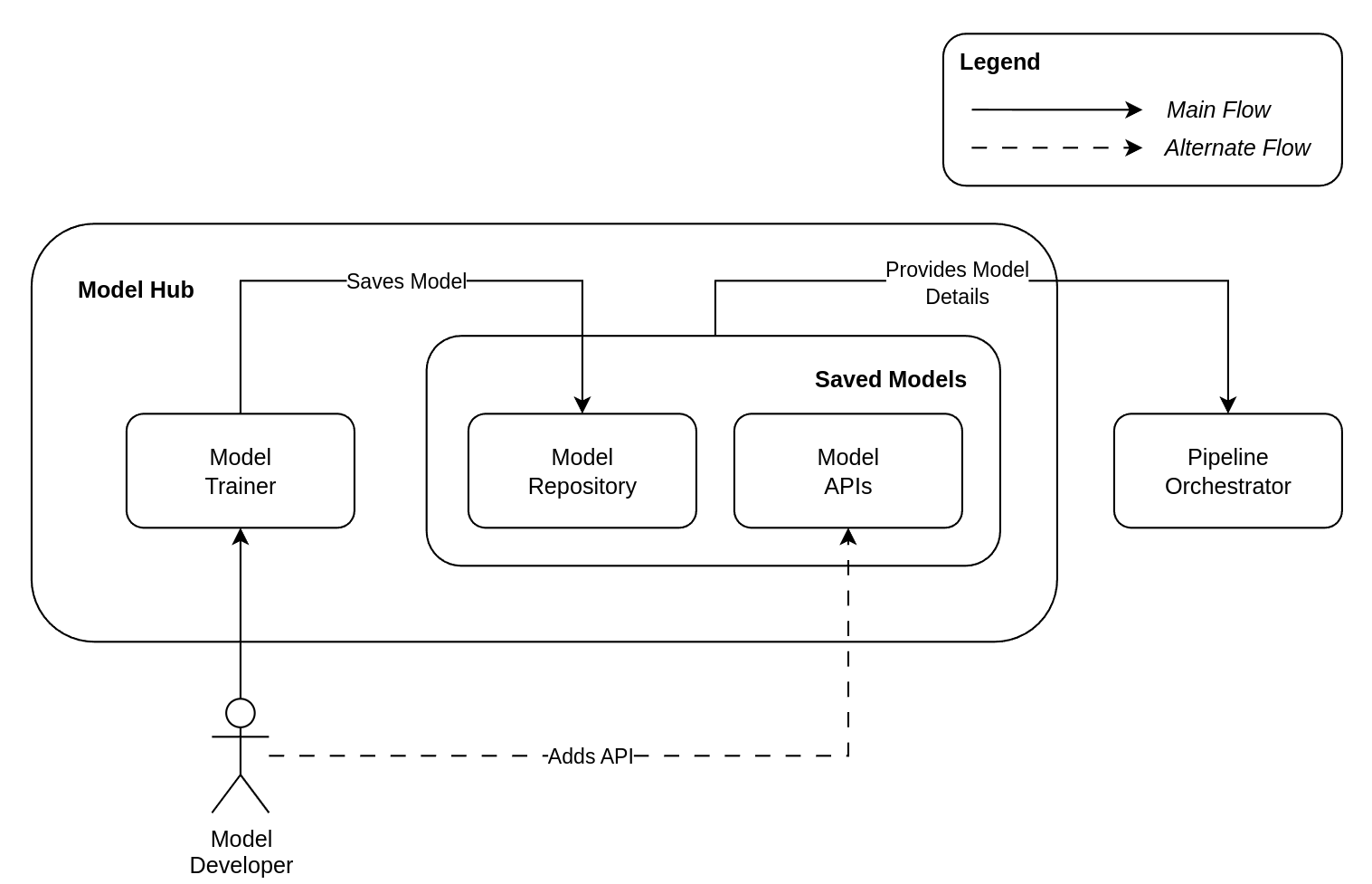}}
    \caption{The Model Hub}
    \label{fig:model_hub}
\end{figure}

The \textit{Model Hub} supplies the system with all the necessary ML models. The \textit{Model Developer} uses this subsystem in one of two ways: either by training an ML model from scratch via the \textit{Model Trainer} and saving the trained model in the \textit{Model Repository}, or by providing an API to an externally deployed model along with a usage mechanism, stored as part of \textit{Model APIs}. Together, the \textit{Model Repository} and \textit{Model APIs} form the \textit{Saved Models} database, which is essential for providing models to the \textit{Pipeline Orchestrator}.
In the \textit{Bhashini Project}, all the necessary models have already been developed and have been made available for use via APIs. We are leveraging these APIs to build our inference pipelines.

\subsection{The Pipeline Orchestrator}
\begin{figure}[htbp]
    \centerline{\includegraphics[width=\columnwidth]{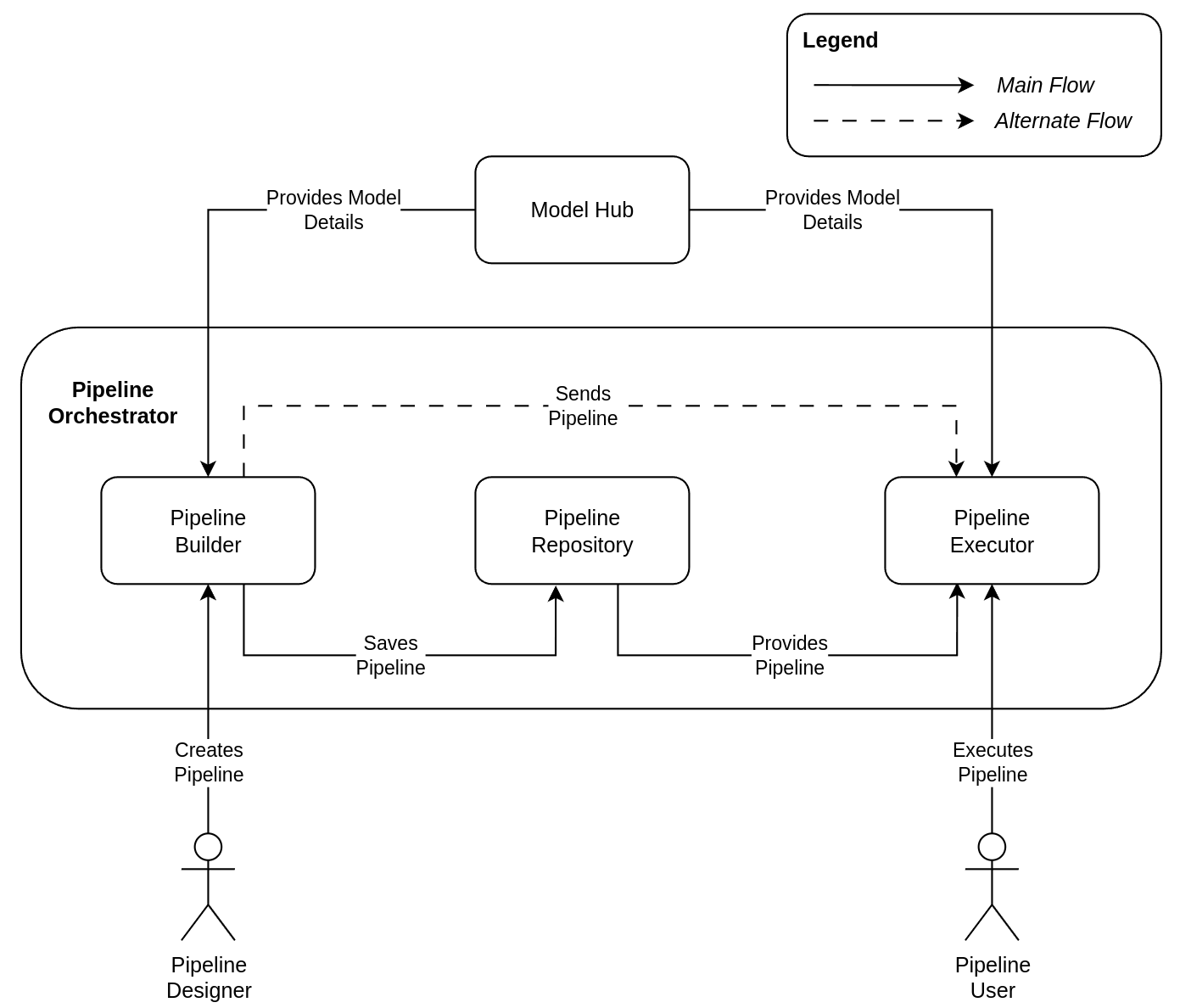}}
    \caption{The Pipeline Orchestrator}
    \label{fig:pipeline_orchestrator}
\end{figure}



The \textit{Pipeline Orchestrator} is responsible for managing all inference-related processes within the system. At the core of this subsystem is the \textit{Pipeline Designer}, who uses the
\textit{Pipeline Builder} component to construct pipelines. A pipeline is essentially a sequence of processing steps, each of which is represented by a node $n_i \in N$ $\forall i \in \{1, ..., k\}$ for a pipeline with $k$ steps. These nodes may include data pre-processing components, various types of machine learning (ML) models, and specialized components known as \textit{adapters}. The sequence in which steps are to be executed is determined by the set of directed edges $E$, where each edge $e_{n_i \rightarrow{} n_j}$ $i \neq j$ denotes that a directed edge from node $n_i$ to node $n_j$ exists in the pipeline.

An \textit{adapter} is a specialized node in the pipeline designed to ensure compatibility between nodes, especially when the output format of one model doesn’t align with the input requirements of the next. They play a critical role when models handle different data types or structures. For instance, an OCR model might output raw text that contains unrecognized or misinterpreted characters when processing an image, whereas a downstream MT model requires clean, structured text as input. The \textit{adapter} acts as an intermediary, cleaning and reformatting the OCR output to meet the MT model's requirements. This bridging function of \textit{adapters} is essential for maintaining smooth data flow and reliable integration among various models within the pipeline, ensuring that each model receives compatible, usable data.


\begin{figure}[htbp]
\centerline{\includegraphics[width=\columnwidth]{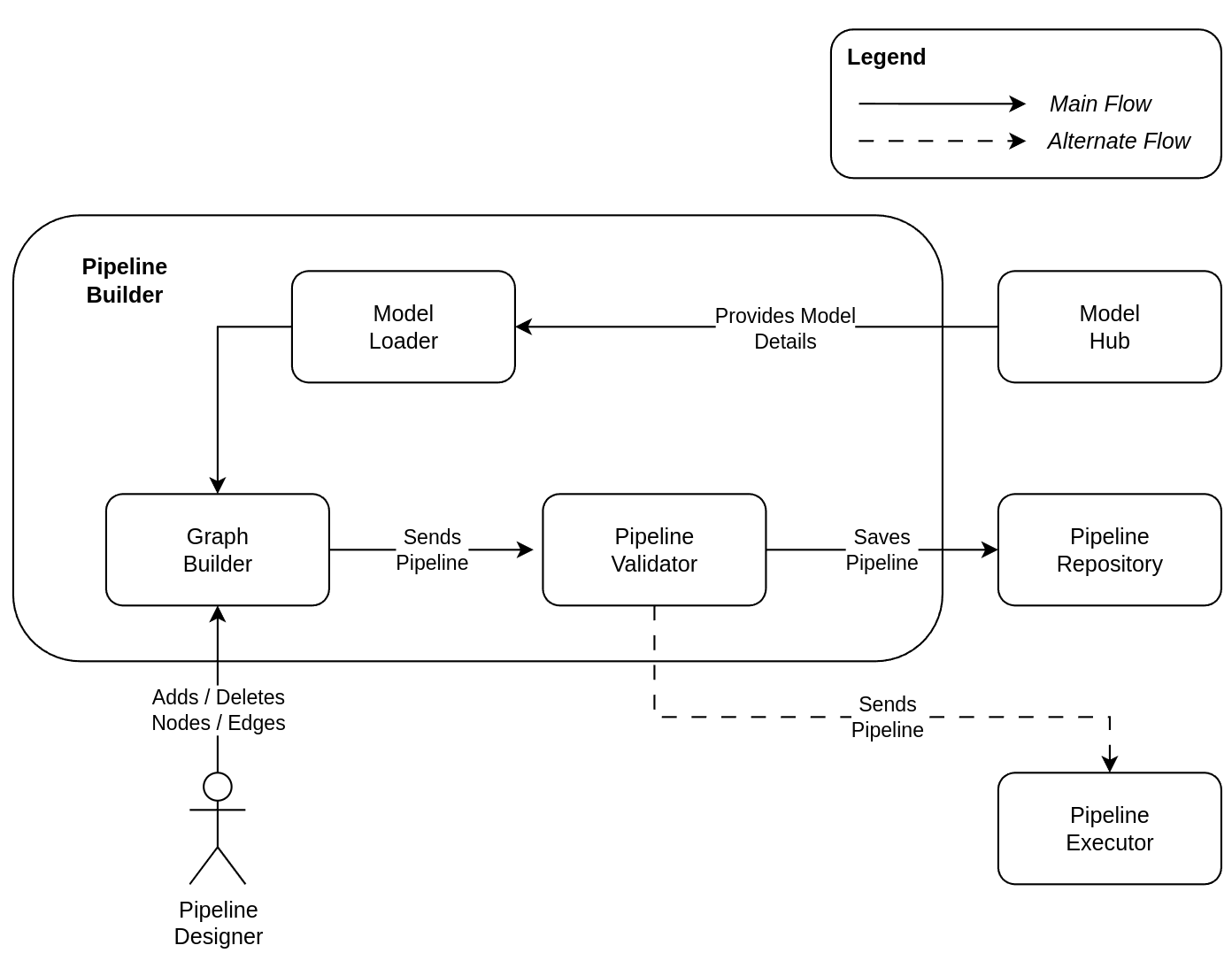}}
    \caption{The Pipeline Builder}
    \label{fig:pipeline_builder}
\end{figure}

The \textit{Pipeline Builder} consists of a \textit{Graph Builder}, which the \textit{Pipeline Designer} uses to create the graph, that is, the set of all nodes $N$ and the set of edges $E$ that define the pipeline’s structure. Model information, sourced from the \textit{Model Hub}, is parsed by the \textit{Model Loader}, which loads the model and makes it available for use as a node within the pipeline.

For instance, in the case of speech-to-text processing as mentioned in Section~\ref{sec:intro}, the \textit{Pipeline Designer} utilizes the \textit{Graph Builder} to construct the pipeline step by step. The process begins by connecting an \textit{input node}, which is a specialized node in the pipeline designed to serve as the entry point for data, (the user can either upload a dataset, or they can provide a link to an external dataset to this node) to an ASR model. The ASR model processes the input audio to generate textual data, which is then linked to the input of a MT model. If a direct MT model translating from the input language to the desired output language is unavailable in the \textit{Model Hub}, multiple MT models can be chained together in sequence. For example, an intermediate translation step might first convert the input text to a bridge language (e.g., English) before translating it into the target language. This chaining capability ensures flexibility in constructing pipelines for complex language translation tasks.

During pipeline construction, validation is managed by the \textit{Pipeline Validator}—a component within the \textit{Pipeline Builder}. Each node $n_i\in N$ has a set of properties $P_{n_i}$ associated with it, that are used to uniquely identify the node itself. We define a rule $r$ to be a construct whose purpose is to check whether some properties of nodes $n_i$ and $n_j$ satisfy some boolean constraint, as defined in Algorithm~\ref{alg:rule}. Note that the \verb|validateRuleConstraint| function is a boolean function that is specific to whichever rule $r$ is utilizing this function.

\begin{algorithm}
\caption{\textit{Rule} evaluation procedure}
\label{alg:rule}
\begin{algorithmic}[1]
\REQUIRE Source Node $n_i$, Destination Node $n_j$
\ENSURE Boolean indicating whether this rule has been satisfied by nodes $n_i$ and $n_j$
 \PROCEDURE{Evaluate}{$n_i, n_j$}
    \IF {property $p_x \in P_{n_i}$ and property $p_y \in P_{n_j}$}
        \RETURN validateRuleConstraint($n_i \rightarrow{} p_x$, $n_j \rightarrow{} p_y$)
    \ELSE
    \RETURN \FALSE
    \ENDIF
 \ENDPROCEDURE
\end{algorithmic}
\end{algorithm}

As defined in Algorithm~\ref{alg:validation}, the Pipeline Validator takes the set of all these rules $r \in R$ as the RuleSet, along with the source node $n_i$ and the destination node $n_j$, and verifies whether all the rules present in the RuleSet are satisfied for a potential edge between these nodes. If they are, then an edge $e_{n_i \rightarrow n_j}$ can exist between them. Else, the user is notified about the invalid edge.

\begin{algorithm}
\caption{\textit{Pipeline Validator}'s edge validation procedure}
\label{alg:validation}
\begin{algorithmic}[1]
\REQUIRE Source Node $n_i$, Destination Node $n_j$, RuleSet $R$
\ENSURE Boolean indicating if an edge $e_{n_i \rightarrow n_j}$ can exist
 \PROCEDURE{CanEdgeExist}{$n_i, n_j, R$}
    \FOR{each rule $r \in R$}
        \IF{not $r.evaluate(n_i, n_j)$}
            \RETURN \FALSE
        \ENDIF
    \ENDFOR
 \RETURN \TRUE
 \ENDPROCEDURE
\end{algorithmic}
\end{algorithm}

In the \textit{Bhashini Project}, the \textit{Pipeline Validator} enforces compatibility rules between model connections to ensure a logical workflow. Specifically, it verifies whether the output of an ASR model is linked to an MT or TTS model’s input, an OCR model's output is connected to an MT or TTS model's input, the output of an MT model is connected to another MT or TTS model’s input, and the output of a TTS model is connected to an ASR model’s input. Furthermore, it also checks whether the model supports the chosen target-source language combination as selected in the nodes.

For example, the \verb|validateRuleConstraint| function, when verifying whether an OCR model can be connected to an MT or TTS model, first checks if the source node ($n_i$) is an OCR model. If this condition is met, the function examines the type of the target node ($n_j$). If the target node is identified as either an MT or TTS model, the function further validates the compatibility of the output language from the OCR model and the input language required by the MT or TTS model. Only if this language pairing is valid for the chosen MT/TTS model, the function returns \verb|true|, indicating the edge is valid. Otherwise, it returns \verb|false|, marking the edge as invalid. If the source node is not an OCR model, this rule does not apply, and the function defaults to returning \verb|true|, as the \textit{Pipeline Validator} ensures that all the rules are satisfied for the given edge. So, any rules that are not applicable must default to \verb|true|. 
\vspace{0.5em}

After a pipeline has been created and validated, the \textit{Pipeline Designer} can request the \textit{Pipeline Builder} to save it in the \textit{Pipeline Repository}, which stores all completed pipelines. Alternatively, the \textit{Pipeline Designer} can choose to test the pipeline without saving by directing the \textit{Pipeline Builder} to send it directly to the \textit{Pipeline Executor}. In this case, the \textit{Pipeline Executor} executes the pipeline without retrieving its details from the Pipeline Repository.

\begin{figure}[htbp]
\centerline{\includegraphics[width=\columnwidth]{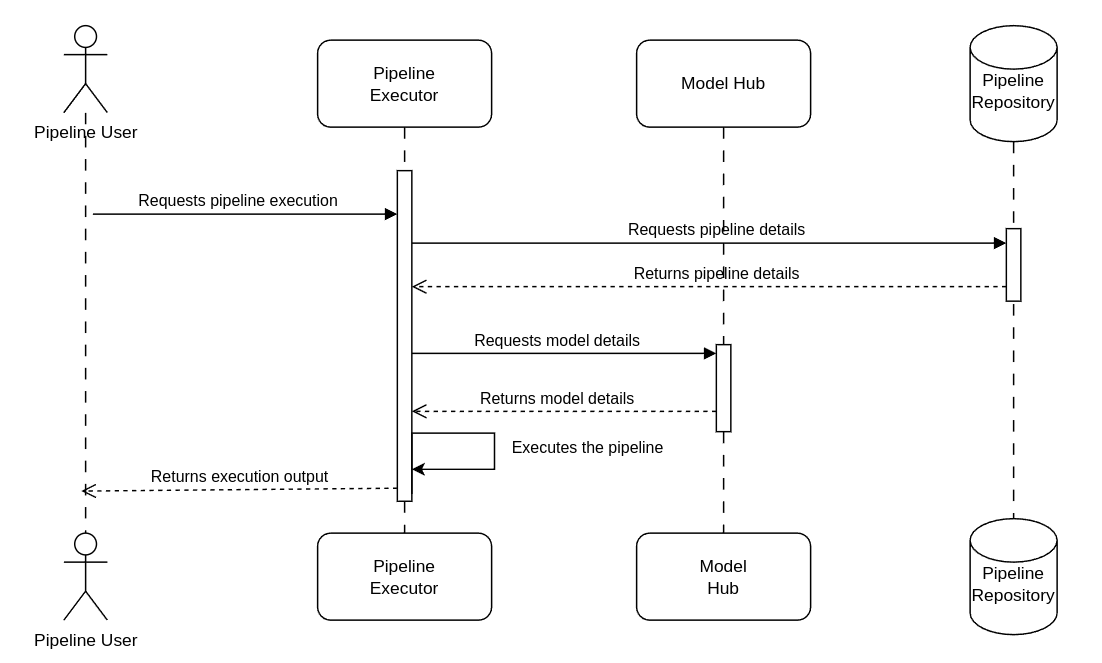}}
    \caption{A sequence diagram of the pipeline execution process}
    \label{fig:flow}
\end{figure}

The \textit{Pipeline User} initiates a pipeline execution request (refer Figure~\ref{fig:flow}), which can be done through various methods, such as sending an API request to a designated endpoint or using a graphical interface. Regardless of the method, the specified pipeline is retrieved from the \textit{Pipeline Repository}, and the model details are loaded from the \textit{Model Hub}.

After all the nodes and edges of the pipeline have been loaded up and are in place, the \textit{Pipeline Executor} plays a central role in handling runtime operations, where it receives the validated pipeline—either from the \textit{Pipeline Repository} or directly from the \textit{Pipeline Builder} as mentioned earlier, for inference tasks. Post the pipeline initiation the \textit{Pipeline Executor} manages data flow across nodes, orchestrating each component's execution based on the defined workflow. During inference, the \textit{Pipeline Executor} processes data inputs through each node, including pre-processing stages, ML models, and \textit{adapters}, ensuring seamless transitions from one stage to the next.
Upon completing the entire pipeline, the final inference output is returned by the \textit{Pipeline Executor} to the \textit{Pipeline User}, providing them with the results of the execution.

In the \textit{Bhashini Project}, the \textit{Inference Zoo} serves as the frontend interface for the \textit{Pipeline Repository}, showcasing all available saved pipelines. Users can browse the \textit{Inference Zoo}, select a desired pipeline for execution, and obtain its corresponding API endpoint (generated at the time the pipeline is saved). By sending input data to this endpoint, the \textit{Pipeline Executor} processes the request, executes the selected pipeline, and returns the output to the user, streamlining the interaction between users and the system.

\section{Preliminary Evaluation}
\label{sec:results}

\begin{figure*}
    \centering
    \includegraphics[width=1\textwidth]{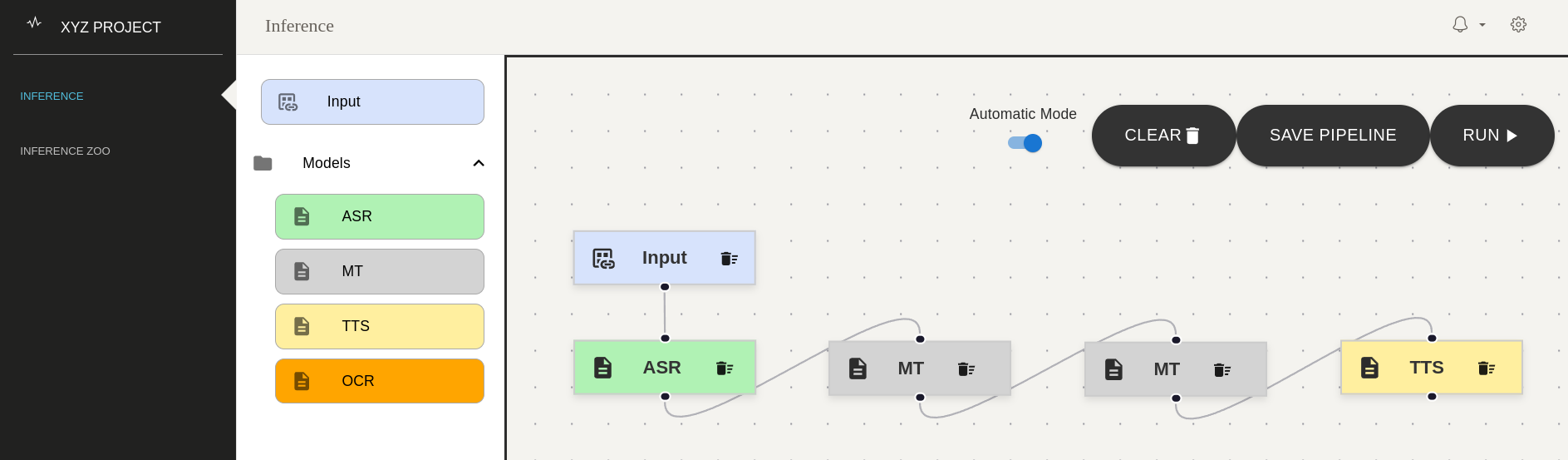}
    \caption{An inference pipeline built using the LoCoML framework}
    \label{fig:XYZ_inference_pipeline}
\end{figure*}

As a preliminary step, experiments were conducted across two distinct machine learning pipeline scenarios. We aimed to measure the additional overhead in running the pipeline that was introduced by the LoCoML platform\footnote{\href{https://anonymous.4open.science/r/XYZ-Project}{https://anonymous.4open.science/r/XYZ-Project}}, on top of the model execution time, in order to demonstrate the overhead's relative impact.

\subsection{Experimental Setup}
We deploy the backend of the LoCoML framework locally in a Docker container, using the Python:3.12-slim base image for the execution environment, on a laptop with the Ryzen 7 5800H CPU that runs at 3.2GHz base speed, and 16GB of DDR4 RAM clocked at 3200MHz. Further, we evaluate LoCoML using two different scenarios:
\begin{enumerate}

\item \textbf{Machine Translation Pipeline}: We tested pipelines containing 1 to 16 MT models.

\item \textbf{Speech Processing Pipeline}: Pipelines cannot consist of only ASR or only TTS nodes because of input-output mismatches—ASR models take audio as input and produce text, while TTS models take text as input and produce audio. Therefore, we tested pipelines with 1 to 8 pairs of ASR and TTS models chained together.
\end{enumerate}

\subsection{Performance Analysis}
\begin{table}[h]
\centering
\begin{subtable}{\columnwidth}
\centering
\caption{Performance Analysis for MT Task}
\resizebox{\columnwidth}{!}{%
\begin{tabular}{|c|c|c|c|}
\hline
\textbf{\# Models} & \textbf{Total Runtime(ms)} & \textbf{Model Runtime(ms)} & \textbf{Additional Overhead(ms)} \\
\hline
1 & 3019.999 & 2967.429 & 52.571 \\
2 & 6537.504 & 6431.963 & 105.541 \\
3 & 8805.692 & 8641.812 & 163.880 \\
4 & 11249.154 & 11045.455 & 203.698 \\
6 & 17639.823 & 17354.786 & 285.037 \\
8 & 23706.186 & 23286.671 & 419.514 \\
12 & 38774.544 & 38170.085 & 604.459 \\
16 & 46882.215 & 46036.834 & 845.381 \\
\hline
\end{tabular}%
}
\end{subtable}

\vspace{0.5cm}

\begin{subtable}{\columnwidth}
\centering
\caption{Performance Analysis for ASR + TTS Task}
\resizebox{\columnwidth}{!}{%
\begin{tabular}{|c|c|c|c|}
\hline
\textbf{\# Pairs} & \textbf{Total Runtime(ms)} & \textbf{Model Runtime(ms)} & \textbf{Additional Overhead(ms)} \\
\hline
1 & 21250.959 & 21144.014 & 106.945 \\
2 & 36543.591 & 36350.016 & 193.576 \\
4 & 71924.239 & 71503.728 & 420.511 \\
6 & 105187.744 & 104544.243 & 643.502 \\
8 & 141696.802 & 140738.753 & 958.050 \\
\hline
\end{tabular}%
}
\end{subtable}
\caption{Performance Comparison of MT and ASR + TTS Tasks}
\label{tab:performance_comparison}
\end{table}

Figure \ref{fig:XYZ_inference_pipeline} demonstrates a pipeline constructed using the drag and drop interface of the LoCoML platform integrated into the \textit{Bhashini project}. Further, Table~\ref{tab:performance_comparison}(a) shows the performance analysis for MT task, while Table~\ref{tab:performance_comparison}(b) shows the performance analysis for ASR + TTS task, and how it scales as the number of models increases.
Our initial results demonstrate that LoCoML's overhead increases linearly with the number of models, scaling from 52.57ms for a single MT model to 845.38ms for 16 models, while remaining negligible at only 1.8\% of the total runtime. This linear overhead growth, coupled with the fact that 98.2\% of the execution time is spent on model inference, indicates that our platform introduces minimal performance impact while effectively managing complex ML pipelines.



\section{Related Work}
\label{sec:related_work}

Recent advances in MDE have highlighted the role of low-code platforms in simplifying ML development and deployment. Naveed et al.~\cite{mde4mlslr} recommend researchers and practitioners to develop low-code platforms for systems with ML components to make ML capabilities more accessible to non-experts, as these platforms can significantly reduce development complexity and time to deployment. Similarly, Iyer et al.~\cite{TrinityCollaboration} introduced Trinity, a no-code platform specifically designed to handle complex spatial datasets, highlighting the versatility and scalability low-code solutions bring to ML applications. In addition, Esposito et al.~\cite{EUDforAISLR} emphasize the importance of user-configurable controls within AI systems, proposing that a balance between automation and manual adjustments can address diverse user needs effectively. Sahay et al.~\cite{sahay2020supporting} provide a detailed survey of various low-code development platforms, identifying key features such as graphical interfaces, interoperability, and scalability as critical for decision-makers evaluating such platforms.

LoCoML builds on these principles, offering a flexible, user-friendly platform that enables non-expert users to construct and customize ML pipelines, facilitating tasks like data preprocessing, model training, and inference without extensive coding knowledge~\cite{bosch2021engineering}. This approach empowers users to iteratively reconfigure workflows, bridging adaptability gaps noted in previous studies and enhancing both accessibility and control~\cite{de2022artificial}~\cite{giray2021software}. Unlike traditional ML systems, LoCoML allows users to adjust pipelines dynamically, aligning with recommendations for integrating user-centric features and configurable controls within ML platforms~\cite{steidl2023pipeline}~\cite{xin2021whither}.

Existing platforms like Azure Machine Learning Designer~\cite{AzureMLDesigner} and AWS SageMaker Pipelines~\cite{AWSSageMaker} provide robust solutions for building and managing machine learning workflows. Azure Machine Learning Designer enables users to create pipelines via a drag-and-drop interface, integrating seamlessly within Azure’s ecosystem. Similarly, AWS SageMaker Pipelines offers a complete suite for creating, deploying, and managing workflows, tightly coupled with AWS-native services. However, these platforms often face challenges when dealing with custom models from external sources, primarily because they are largely designed to operate within their respective ecosystems~\cite{AzureMLCustomModel} \cite{AWSSageMakerCustomModel}. It is too tedious to have to port custom models into the specific input-output constraints as prescribed by these platforms. In contrast, our framework addresses this gap by providing a unified and adaptable solution capable of accommodating diverse, partner-contributed models, ensuring compatibility and seamless integration—capabilities that are currently absent in these existing platforms.

\section{Conclusion and Future Work}
\label{sec:conclusion}

To conclude, we have introduced LoCoML, a low code framework designed to streamline the development of ML inference pipelines. 
LoCoML has been successfully integrated into the \textit{Bhashini Project} with a drag-and-drop interface to create pipelines, where it operates in a real-world production environment, supporting users in building and managing inference pipelines efficiently. The framework's simple interface allows users, including those without extensive coding skills, to connect and control various ML models seamlessly. Our evaluation across multiple scenarios, including TTS, MT, and ASR, indicate that LoCoML has significantly simplified the process of constructing complex, multimodel workflows, making ML pipeline development more accessible and practical for a diverse range of users. 


Looking ahead, we aim to expand LoCoML's capabilities in response to evolving requirements within the \textit{Bhashini Project}. Stakeholders of this project are exploring the potential of extending the framework to support model training, thus creating an end-to-end solution covering both training and inference within the same pipeline. Additionally, we plan to conduct further studies to assess LoCoML's effectiveness in terms of user experience, usability, and performance. These future enhancements will ensure that LoCoML continues to evolve as a versatile and robust framework, meeting the growing demands of ML practitioners and researchers within the \textit{Bhashini Project} and beyond.

\section*{Acknowledgment}
The authors acknowledge the anonymous reviewers for their valuable feedback. The authors thank 
Ayush Agrawal,
Harshit Karwal,
Mukta Chanda,
Rohan Chowdary,
Shashwat Dash,
Siddharth Mavani, and
Supreeeth S Karan, for their assistance in developing the code artifacts necessary to build this framework.
The authors would also like to acknowledge \textit{Bhashini Engineering Unit}~\footnote{https://bhashini.gov.in/sahyogi/anushandhan-mitra/15} team for the support.



\bibliographystyle{ieeetr}
\bibliography{references}

\end{document}